\documentclass[journal=jctcce,manuscript=article]{achemso}

\author{Michiaki Arita}
\affiliation[Tokyo University of Science]
{Faculty of Science and Technology, Tokyo University of Science,
2641 Yamasaki, Noda, Chiba 278-8510, Japan}
\alsoaffiliation[CMSU, NIMS]
{Computational Materials Science Unit (CMSU), National Institute for Materials Science (NIMS),
1-1 Namiki, Tsukuba, Ibaraki 305-0044, Japan}

\author{David R. Bowler}
\affiliation[Physics and Astronomy, UCL]
{Department of Physics and Astronomy, University College London (UCL),
Gower Street, London WC1E 6BT, UK}
\alsoaffiliation[LCN, UCL]
{London Centre for Nanotechnology(LCN), University College London (UCL),
17-19 Gordon Street, London WC1H 0AH, UK}
\alsoaffiliation[MANA, NIMS]
{International Center for Materials Nanoarchitectonics (MANA), National Institute for Materials Science (NIMS),
1-1 Namiki, Tsukuba, Ibaraki 305-0044, Japan}

\author{Tsuyoshi Miyazaki}
\email{MIYAZAKI.Tsuyoshi@nims.go.jp}
\affiliation[CMSU, NIMS]
{Computational Materials Science Unit (CMSU), National Institute for Materials Science (NIMS),
1-1 Namiki, Tsukuba, Ibaraki 305-0044, Japan}
\alsoaffiliation[Tokyo University of Science]
{Faculty of Science and Technology, Tokyo University of Science,
2641 Yamasaki, Noda, Chiba 278-8510, Japan}

\title{Stable and Efficient Linear Scaling First-Principles Molecular
  Dynamics for 10,000+ atoms}

\begin{document}

\begin{tocentry}

Some journals require a graphical entry for the Table of Contents.
This should be laid out ``print ready'' so that the sizing of the
text is correct.

Inside the \texttt{tocentry} environment, the font used is Helvetica
8\,pt, as required by \emph{Journal of the American Chemical
Society}.

The surrounding frame is 9\,cm by 3.5\,cm, which is the maximum
permitted for  \emph{Journal of the American Chemical Society}
graphical table of content entries. The box will not resize if the
content is too big: instead it will overflow the edge of the box.

This box and the associated title will always be printed on a
separate page at the end of the document.

\end{tocentry}

\begin{abstract}
The recent progress of linear-scaling or O($N$) methods in the density functional theory (DFT) 
is remarkable. 
Given this, we might expect that first-principles molecular dynamics (FPMD) simulations based on DFT 
could treat more realistic and complex systems using the O($N$) technique.
However, very few examples of O($N$) FPMD simulations exist so far and the information 
for the accuracy or reliability of the simulations is very limited.
In this paper, we show that efficient and robust O($N$) FPMD simulations are now possible by 
the combination of the extended Lagrangian Born-Oppenheimer molecular dynamics method, 
which was recently proposed by Niklasson {\it et al} ({\it Phys. Rev. Lett.} 100, 123004 (2008)), 
and the density matrix method as an O($N$) technique.
Using our linear-scaling DFT code {\sc Conquest}, 
we investigate the reliable calculation conditions for the accurate O($N$) FPMD 
and 
demonstrate that we are now able to do practical, reliable self-consistent FPMD simulation of 
a very large system containing 32,768 atoms.
\end{abstract}

\section{Introduction}
\label{sec:introduction}

First-principles molecular dynamics (FPMD) based on density functional theory (DFT) is
a well-established and highly-successful tool for studying reactions or processes of materials at the atomic scale.
Due to the increase of the computer power,
the variety and complexity of the materials and phenomena investigated by FPMD simulations have been growing.
However, the size of the systems modelled with FPMD simulations have
remained limited to systems 
of a few hundred atoms in most cases, because the computational cost of standard DFT methods grow rapidly, 
proportional to the cube of the number of atoms $N$ in the system. 
There are many demands to enlarge the system size in diverse fields,
including computational physics, chemistry, 
materials science, biology and so on.
Examples of these problems include chemical reactions at liquid/solid interfaces, 
various processes in complex biological systems, or the growth mechanism of the nano-structured 
materials at atomic scale; in all these cases, we have to treat
systems containing many thousands or tens of thousands of atoms.
In this respect, recent advances in developing linear-scaling or O($N$) methods are encouraging\cite{Bowler:2012zt}.
There have been several demonstrations that efficient and reliable linear-scaling DFT calculations
are now available to calculate the electronic structure, total energy and atomic forces of very
large systems, including up to millions of atoms\cite{Bowler:2010uq,Arita.2014.JASSE}.

Although it is now possible to calculate the total energy and atomic forces of very large systems
using O($N$) DFT methods\cite{2002.soler.2475, 2004.miyazaki.6186, Hine.2011.PRB, Osei-Kuffuor.2014.PRL}, 
this does not guarantee that stable, efficient and accurate FPMD
simulations are possible in practice\cite{Bowler.2014.arXiv}.
With conventional DFT methods, there are two widely-used methods to achieve efficient FPMD simulations.
The first method is known as Car-Parrinello MD (CPMD)\cite{Car.1985.PRL}, where the propagation of
electronic structure and the atomic positions are treated simultaneously by introducing
a fictitious mass for the electronic degree of freedom.
This method is efficient, but the accuracy of the CPMD simulations can
depend on the choice of the fictitious mass.
By contrast, Born-Oppenheimer MD (BOMD) deals with the electronic and nuclear
problems separately. In this method, the electronic structure in the ground state is calculated at each
set of atomic positions,  usually by the optimization of the Kohn-Sham orbitals using an iterative method\cite{Payne.1992.RMP}.
The BOMD method thus needs more CPU time than CPMD for each MD step, but
the method is robust and stable, and allows us to adopt a longer time step.
As a result, the CPU time needed for a whole MD simulation is tractable because the number of force
calculations is smaller (the cost of calculating atomic forces is usually much higher than
the update of the Kohn-Sham orbitals).
The main problem facing this method, however, is that both the stability and the reliability of the MD depend on
the accuracy of the calculated forces.
If the optimization of the Kohn-Sham orbitals or the convergence of
SCF is not sufficiently accurate,
the method suffers from a systematic energy drift in micro-canonical MD simulations,
due to the violation of time-reversal symmetry.
This can also result in poor reliability for canonical MD simulations.
Recently this problem has been solved using the extended Lagrangian Born-Oppenheimer MD (XL-BOMD) proposed
by Niklasson et al.\cite{2006.Niklasson.PRL,2008.Niklasson.PRL,2011.Niklasson.JCP}, 
whose Lagrangian includes auxiliary electronic degrees of freedom as
dynamical variables to recover time-reversal symmetry in BOMD simulations.
They have shown that BOMD simulations with long-term conservation of total energy are possible.

It is reasonable to expect that this recent progress with the XL-BOMD
method could be used with O($N$) DFT techniques.
However, since linear-scaling methods rely on the locality of the electronic structure and
need to introduce approximations to utilize this locality, the accuracy of O($N$) method may change during
the MD simulations, and it is not clear how accurate or stable MD can
be with O($N$) DFT methods.
Although there have been already a few reports showing the examples of O($N$) 
FPMD\cite{Ohwaki.2012.JCP,Ohwaki.2014.JCP,Shimojo.2014.JCP,Tsuchida:2008ai}
or O($N$) self-consistent tight-binding\cite{2012.Cawkwell.JCP} simulations, 
they use different linear scaling techniques and
the information about the accuracy, stability or efficiency of the simulations is rather limited.
In this paper, we report the stability and accuracy of the XL-BOMD simulations with
a density matrix minimization (DMM) technique\cite{li1993,1995.hernandez.10157}, which is one of the most 
common O($N$) DFT methods.
The accuracy of the forces calculated by the DMM method is high because the method satisfies the
variational principle\cite{2004.miyazaki.6186}.
To our knowledge, the combination of DMM and XL-BOMD methods has not been investigated so far.
Since the original XL-BOMD methods assume the orthogonality of the basis functions, the formulation
of the method for non-orthogonal basis functions is also presented.
We show that the XL-BOMD method can be introduced to our linear-scaling DFT code 
{\sc Conquest}\cite{1996.hernandez.7147,2002.bowler.2781,2006.bowler.989}, and
report that robust, accurate and efficient FPMD is possible with the DMM method.
The reliable calculation conditions for the accurate O($N$) FPMD are also investigated.
In the end, we demonstrate that it is now possible to do actual FPMD simulations of a 32,768-atom system.

The rest of the paper is organized as follows. In the next section, we show the methods and algorithms
which are used or introduced in this work. In Sec. 3, we report the results of the test calculations,
investigation of the reliable calculation conditions, and some examples of the FPMD simulations, including
that of 32,768-atom system. Finally, concluding remarks are given in Sec. 4.

\section{Methods and Algorithms}
\label{sec:methods-algorithms}

\def\CQ{\textsc{Conquest}}

The implementation of molecular dynamics and all calculations were performed with \CQ, 
which is a linear scaling, pseudopotential DFT code\cite{1996.hernandez.7147,2002.bowler.2781,2006.bowler.989}.
In order to achieve linear-scaling both in computational time and
memory consumption, the code works with the density matrix $\rho$
rather than the wavefunctions, because of its spatial locality:
\begin{equation}
\rho\left(\mbox{\boldmath $r$}, \mbox{\boldmath $r$}^{\prime} \right)
\rightarrow 0 \mbox{ \textrm{as} }
|\mbox{\boldmath $r$} - \mbox{\boldmath $r$}^{\prime}| \rightarrow
\infty.
\end{equation}
\CQ\ assumes that the density matrix can be described in a separable form:
\begin{eqnarray}
        \rho(\mbox{\boldmath $r$}, \mbox{\boldmath $r$}^{\prime}) =
        \sum_{i\alpha, j\beta} \phi_{i\alpha}(\mbox{\boldmath $r$})K_{i\alpha,j\beta}\phi_{j\beta}(\mbox{\boldmath $r$}^{\prime}),
\end{eqnarray}
where $\phi_{i\alpha}$ is a non-orthogonal localised orbital called a \textit{support function},
which is only non-zero within a sphere centred at atom $i$,
and $\alpha$ denotes the support functions of a given atom.

Support functions are represented in terms of localised basis
functions; in \CQ\ we can use systematically improvable basis
functions (B-splines\cite{1997.hernandez.13485}) or pseudoatomic-orbitals
(PAOs)\cite{2002.soler.2475,2008.torralba.294206}.
$K_{i\alpha,j\beta}$ is the density matrix represented in terms of the support functions.
We find the electronic ground state by minimising the total energy
with respect to the density matrix, while imposing the correct
electron number and weak idempotency.
The complete minimisation combines two algorithms\cite{Bowler:1999if}:
McWeeny's iterative purification\cite{McWeeny:1960zp,Palser:1998fd}; and the auxiliary
density matrix (ADM) method proposed by Li, Nunes and Vanderbilt\cite{li1993}.
An initial density matrix is generated starting from the Hamiltonian
(scaled to have the right trace and eigenvalue spectrum),
using the extension of McWeeny's approach for the canonical case\cite{Palser:1998fd}.

In the ADM method, weak idempotenty is imposed by expressing the density
matrix $K$ in terms of an  auxiliary density matrix $L$,  and the overlap $S$
between support functions, $S_{i\alpha,j\beta} = < {\phi_{i\alpha}|\phi_{j\beta}} >$, as
\begin{eqnarray}
        K = 3LSL-2LSLSL
 \label{eq:purification}
\end{eqnarray}
The spatial truncation on the density matrix is actually 
imposed 
on the matrix $L$, such that $L_{i\alpha,j\beta} = 0$ once
$| \mbox{\boldmath $R$}_i - \mbox{\boldmath $R$}_j | \ge R_L $; this
is referred as \textit{L-range} in the rest of this report.

The McWeeny's procedure is applied until the energy rises (a sign of
reaching truncation error\cite{Palser:1998fd}).  The $L$ matrix is then passed into the ADM,
which is a variational approach, using a modified Pulay/Broyden algorithm.
This part of the minimisation is referred to as DMM (density matrix minimisation).
%
\\

In this paper, we represent each support function $\phi_{i, \alpha}(r)$
in terms of one PAO (the approach given here is easily extended to the case where we also minimise
the energy with respect to the PAOs, which will be presented in a future publication).
In running BOMD simulations with the DMM method, we have two ways to optimize the electronic structure.
One is to initialise the density matrix using McWeeny's procedure at every atom movement.
The initial charge density is made from the superposition of the atomic charge density.
Since the Hamiltonian and overlap matrices are recalculated after the movement of atoms,
the initial matrix $L$ and subsequent optimised $L$ matrix will maintain
time-reversal symmetry; we hence expect that BOMD simulations should
conserve energy and be stable.
The other method is to utilize the auxiliary density matrix found from
the previous MD step to start the present DMM.
In this case, the initial charge density is made from the initial 
density matrix $L$ prepared for the present DMM, and is updated with the $L$
matrix simultaneously during the DMM.
This method will be more efficient, but will break time reversal symmetry, potentially
leading to unphysical energy drift.

To maintain time-reversal symmetry while keeping the efficiency of
density matrix re-use, we will use the extended Lagrangian scheme proposed by Niklasson et al.\cite{2008.Niklasson.PRL}
We introduce an auxiliary degree of freedom to the Born-Oppenheimer
Lagrangian $\mathcal{L}^{\rm XBO}$, $X$.  This is associated with $LS$,
rather than $L$ to maintain the correct non-orthogonal metric\cite{footnote1}.
\begin{eqnarray}
        {\cal L}^{\rm XBO}\left( X, \dot{X}, \mbox{\boldmath $R$}, \dot{\mbox{\boldmath $R$}} \right) =
        {\cal L}^{\rm BO}\left( \mbox{\boldmath $R$}, \dot{\mbox{\boldmath $R$}} \right) +
        \frac{\mu}{2} {\rm Tr}\left[ \dot{X}^2 \right] -
        \frac{\mu \omega^2}{2} {\rm Tr}\left[ \left( LS  - X \right)^2 \right]
  \label{eq:XLBOMD}
\end{eqnarray}
$X$ is a sparse matrix with the range of the matrix $LS$, $\mu$ the fictitious electronic mass,
and $\omega$ is the curvature of the electronic harmonic potential.
As in the original XL-BOMD method, if we take the limit $\mu \rightarrow 0$,
$\mathcal{L}^{\rm XBO}$ becomes $\mathcal{L}^{\rm BO}$ and we have equations of motion for nuclear positions and $X$.
If we apply the time-reversible Verlet scheme to calculate $X$ using the equation of motion, we have
\begin{eqnarray}
  X(t+\delta t) = 2X(t) - X(t-\delta t) + \delta t^2 \omega^2 (L(t)S(t)- X(t))
  \label{eq:TRBOMD}
\end{eqnarray}
which shows that $X(t)$ is time reversible and evolves in a harmonic potential
centered around the ground-state $L(t)S(t)$.
This also implies that a good initial guess for the $L$-matrix, which will
obey time reversal symmetry, can be calculated by multiplying $X$ and
$S^{-1}$ (in \CQ, the sparse approximate inverse $S$ is computed using Hotelling's method\cite{Press:1992dp}).

Even though the time-reversibility is maintained during propagation,
the $X$-matrix tends to move away from the harmonic centre
over time.  As a result, the number of iterations to reach the
ground state at each MD step gradually increases in the course of a simulation.
To remove accumulated numerical errors, we add a dissipative force to
the propagation of $X$  
following Ref.~\cite{2009.Niklasson.JCP}, 
\begin{eqnarray}
  X(t+\delta t) = 2X(t) - X(t-\delta t) + \kappa (L(t)S(t)- X(t))
                                        + \alpha \sum_{m=0}^M c_m X(t-m \delta t).
  \label{eq:dissipation}
\end{eqnarray}
Here, the parameters $\kappa$, $\alpha$ and $c_m$ are determined so that the dissipation term
does not significantly break the time-reversal symmetry; we use the values in Ref.~\cite{2009.Niklasson.JCP}. 
The amount of dissipation is controlled with a single parameter $M$,
which determines the order of the polynomial, and we use $M=5$ in this paper.
With the dissipation term, we can keep the kinetic energy of $\dot{X}$ small and 
thus the $X(t)$ is kept close to the present ground-state $L(t)S(t)$.
As a result, we expect that the number of DMM steps would be reduced.
The effect of the dissipation term in the practical calculations is reported and 
discussed in Sec.~\ref{sec:effect-schem-line}.


\section{Results}
\label{sec:results}

We will present our tests of the implementation of O($N$) DFT in different stages:
first, considering approaches that permit energy conservation and their efficiencies;
second, the effect of density matrix range and minimisation tolerance;
and finally presenting practical calculations.

\subsection{Effective schemes for linear-scaling BOMD}
\label{sec:effect-schem-line}

First, we explore methods to achieve energy conservation with O($N$) DFT and BOMD.
We monitor the Born-Oppenheimer total energy $E_{\rm BO}$, which is defined as a sum of
the ionic kinetic energy $T$ and the DFT total energy $V_{\rm BO}$.
In the micro-canonical ensemble, $E_{\rm BO}$ should be constant and so is a good indicator
to judge whether a simulation is accurately carried out.
In what follows, $R_{L}$ is set to 16 bohr (the effect of changing this is explored
below in Sec.~\ref{sec:calc-cond-bmmathc}).
A single-$\zeta$ (SZ) PAO is used to represent support functions\cite{footnote2}, 
and the local density approximation (LDA) is adopted to the exchange-correlation functional with the
parametrization by Perdew and Zunger\cite{Perdew.1981.PRB}. MD simulations are conducted via the
velocity-Verlet integrator\cite{PhysRev.159.98} with a time step being 0.5 femtoseconds (fs) in a micro-canonical ensemble;
the initial velocities are given randomly so that the Maxwell-Boltzmann distribution
holds with 300 K.
The same conditions are set in other simulations in this report unless otherwise specified.

\begin{figure}[H]
        \begin{center}
                \includegraphics[width=9cm,height=7.5cm]{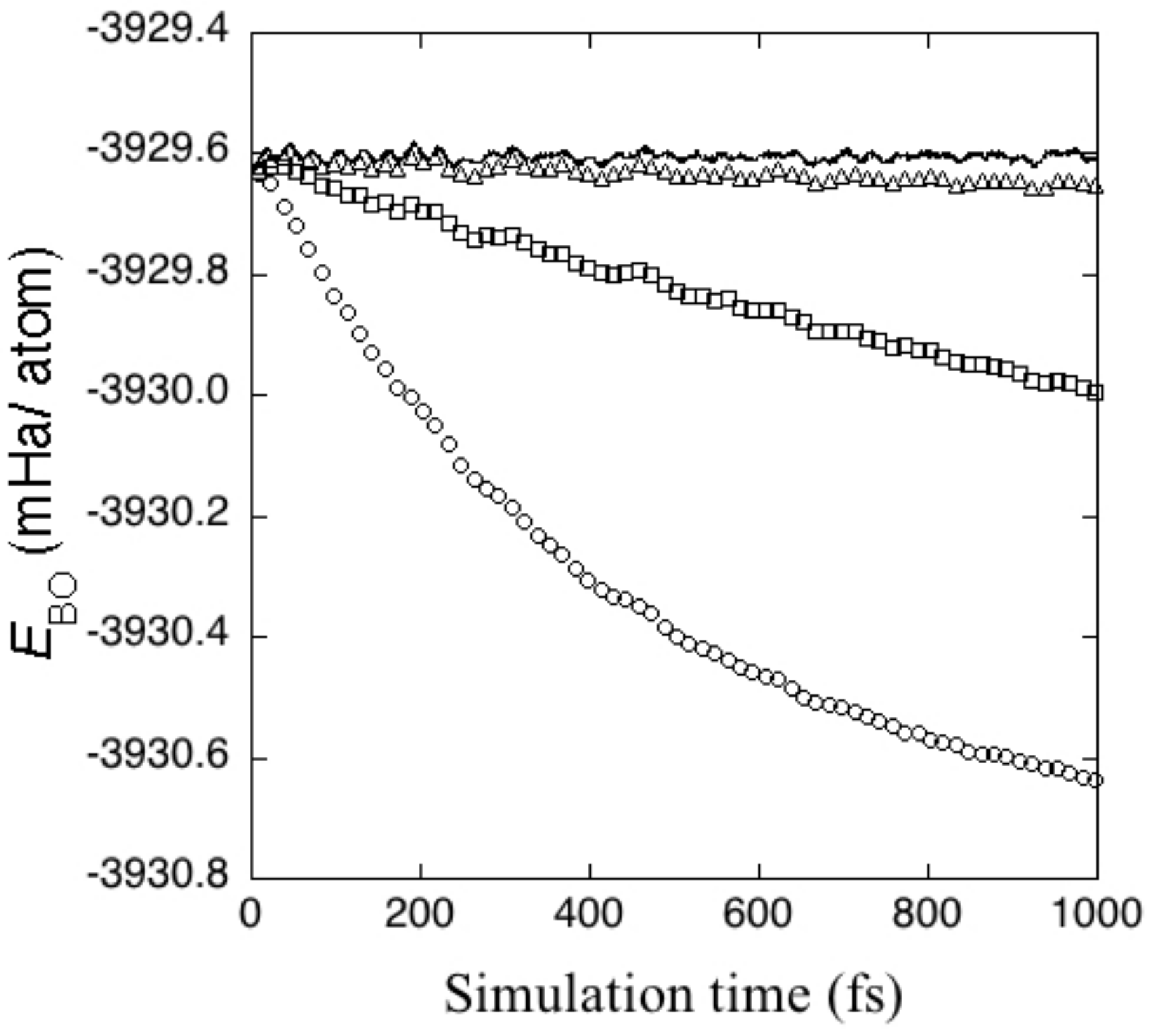}
                \caption{Born-Oppenheimer total energy profiles obtained with McWeeny initialisation at every step (solid line) and by reusing the L-matrix from the previous step for different tolerances (symbols).  Symbols stand for tolerance $\epsilon_L$ taking values 1.6$\times10^{-5}$ (circle), 1.6$\times10^{-7}$ (square) and 1.6$\times10^{-9}$ (triangle).}
                \label{fig:Eprofiles1}
        \end{center}
\end{figure}

For a first run, we perform BOMD on a 64-atom crystalline silicon with density-matrix
initialisation by McWeeny step, so that the density matrix is initialised from the
Hamiltonian at each MD step.
The density matrix is updated until the residual in the DMM step,
$\frac{1}{N} {\rm Tr}(\sigma S^{-1}
\sigma S^{-1})$ with $\sigma = \delta E^{\prime}/\delta L$,
becomes smaller than a given tolerance $\epsilon_L$.
Here, $E^{\prime}$ is defined as $V_{\rm BO}-\mu N_{\rm el}$, the
number of electrons is $N_{\rm el}$ and 
the Lagrange multiplier $\mu$ is used to keep the electron number fixed.
The profile of the Born-Oppenheimer energy $E_{\rm BO}$ is shown by a solid line in Fig.~\ref{fig:Eprofiles1}, and it is clear
that the total energy is conserved even for a low DMM tolerance, $\epsilon_L=1.6\times10^{-5}$.
This simulation is, however, time-consuming since the electronic structure is calculated
from scratch at every atom movement.

We now consider the re-use of the optimised $L$-matrix from the previous MD step.
When the optimised $L$-matrix from the previous step is used to initialise the density matrix,
the \textsc{Cpu} time is significantly reduced.
However, as well as this improved efficiency, we observe a significant and undesired energy drift,
as shown in Fig.~\ref{fig:Eprofiles1} with symbols.
Here, we consider various values for the tolerance applied to the density matrix minimisation, $\epsilon_L$.
The energy drift can be reduced by using a tighter $\epsilon_L$, but tighter tolerance
rapidly increases the number of DMM iterations and the computational cost.

\begin{figure}[H]
        \begin{center}
                \includegraphics[width=9cm,height=7.5cm]{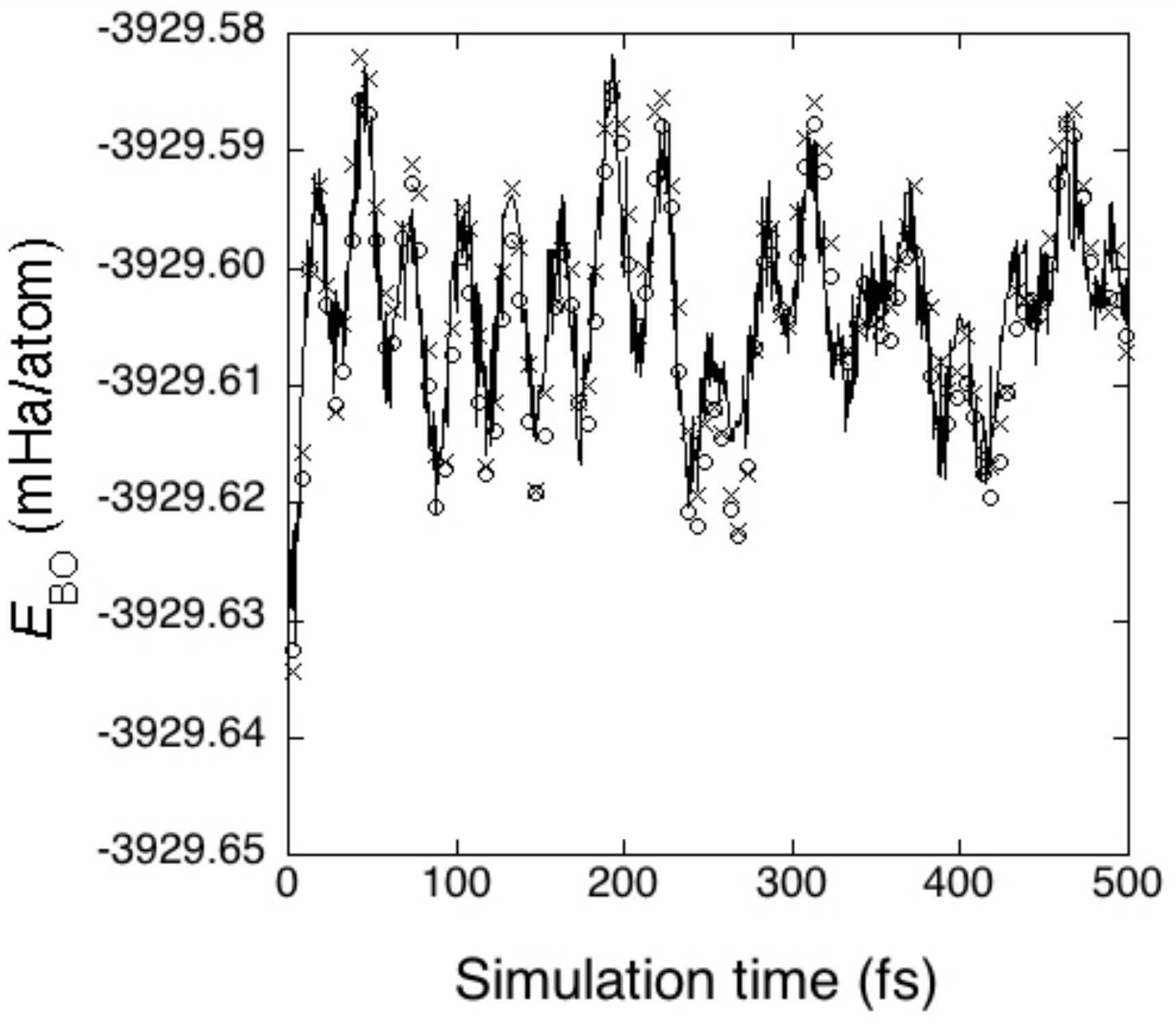}
                \caption{Born-Oppenheimer total energy profiles with tolerance $\epsilon_L$ 1.6$\times10^{-5}$, and different initialisations: McWeeny initialisation at each MD step (solid line); XL-BOMD with dissipative force (circles);  and XL-BOMD without dissipative force (crosses).}
                \label{fig:Eprofiles2}
        \end{center}
\end{figure}

In order to solve the problem of both accuracy and efficiency, we carry out XL-BOMD with
a DMM tolerance, $\epsilon_L=1.6\times10^{-5}$.
Its $E_{\rm BO}$ profile (crosses in Fig.~\ref{fig:Eprofiles2}) agrees extremely well with 
that of BOMD with density-matrix initialisation by means of McWeeny step 
(solid line in Fig.~\ref{fig:Eprofiles2}), implying that 
XL-BOMD yields a well conserved energy with lower computational requirements.
However, there is a problem in the efficiency with the method, as explained in Sec. 2.
After a long simulation time, the number of DMM iterations required to find the ground state density matrix grows.
This growing cost can be suppressed by applying a dissipative force, and the resulting $E_{\rm BO}$
profile, circles in Fig. ~\ref{fig:Eprofiles2}, is still very close to the profile without dissipation.
A comparison of the required number of DMM iterations at each step for these different approaches
is shown in Table.~\ref{table:DMMcost}.
The reduction of computational cost by using XL-BOMD with dissipation is significant.
Besides the smaller number of DMM iterations, the method does not require McWeeny steps which
typically include a few tens of matrix multiplications for purification (as in Eq.~\ref{eq:purification}).

\begin{table}[H]
        \begin{center}
                \begin{tabular}{l r r r r} \\
                        \hline \hline
                        type of run & dissipation & average & max & min \\
                        \hline
                        BOMD & N & 10 & 12 & 8 \\
                        XL-BOMD & N & 6.7 & 8 & 2\\
                        XL-BOMD & Y & 3.9 & 4 & 2 \\
                        \hline\hline
                \end{tabular}
                \caption{DMM iterations required for each MD step; BOMD with density-matrix initialisation by McWeeny step, and XL-BOMD in the presence (Y) and absence (N) of a dissipative force.  Averages taken over 1,000 steps.}
                \label{table:DMMcost}
        \end{center}
\end{table}

\subsection{Effects of density matrix tolerance and range on XL-BOMD}
\label{sec:calc-cond-bmmathc}

In the last section, we showed that XL-BOMD simulation with DMM method was robust and efficient
using one typical set of calculation conditions.
We now investigate how MD results are affected by the parameters which control the accuracy
and efficiency of linear-scaling DFT calculations with DMM.
There are two key parameters: the range applied to the auxiliary density matrix ($L$-range or $R_L$);
and the tolerance applied to the minimisation, $\epsilon_L$.
In both tests, XL-BOMD is performed on a 64-atom crystalline silicon with a dissipative force.

\subsubsection{\bf{$R_L$}-dependence}

The key approximation made in linear scaling is to localize the
density matrix, and it is vital to examine how the conservation of energy, $E_{BO}$, is affected by the range,
the parameter $R_L$.  We have tested three values of $R_L$: 13, 16 and 20 bohr, each with a strict
DMM tolerance, $\epsilon_L = 1.6\times10^{-7}$.
Simulations are performed for 2ps (or 4,000 steps).

\begin{figure}[H]
        \begin{center}
                \includegraphics[width=7cm,height=10cm]{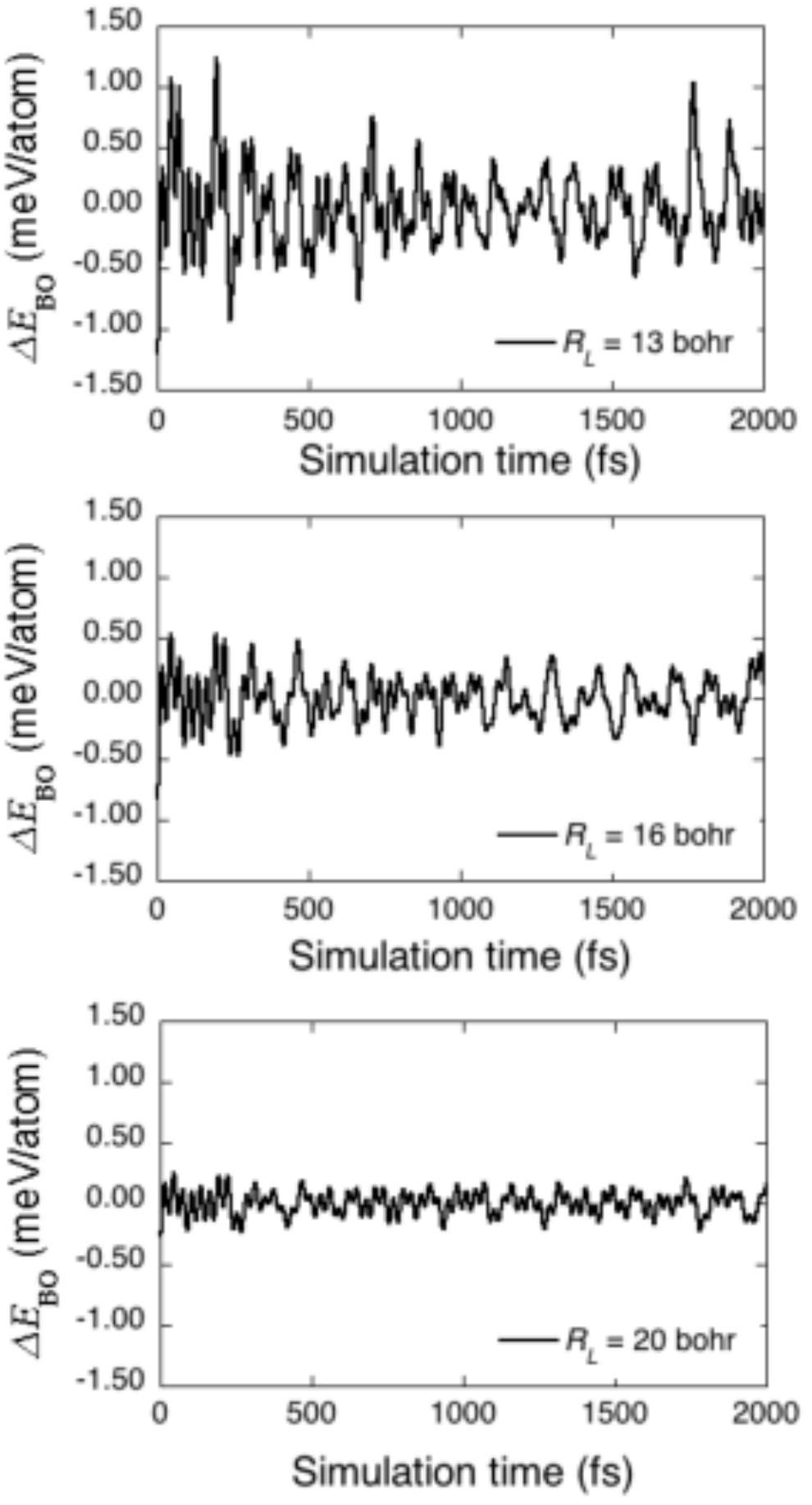}
                \caption{The instantaneous energy fluctuations calculated by three $R_L$ values; 13 bohr (top), 16 bohr (middle) and 20 bohr (bottom).}
                \label{fig:RLdep}
        \end{center}
\end{figure}

The time-averaged BO total energies are very different for the three ranges (-106.830 eV/atom,
-106.930 eV/atom and -106.987 eV/atom) simply because the potential energy is lower for longer ranges.
We plot the instantaneous energy fluctuations for the three ranges in Fig.~\ref{fig:RLdep};
the time-averaged absolute fluctuations, $\left< \left| \Delta E_{BO} \right| \right>$,
are 0.24 meV/atom, 0.15 meV/atom and 0.075 meV/atom.
These fluctuations become smaller as the range increases, and no energy drift is seen for any of the ranges.

\begin{figure}[H]
        \begin{center}
                \includegraphics[width=10cm,height=9cm]{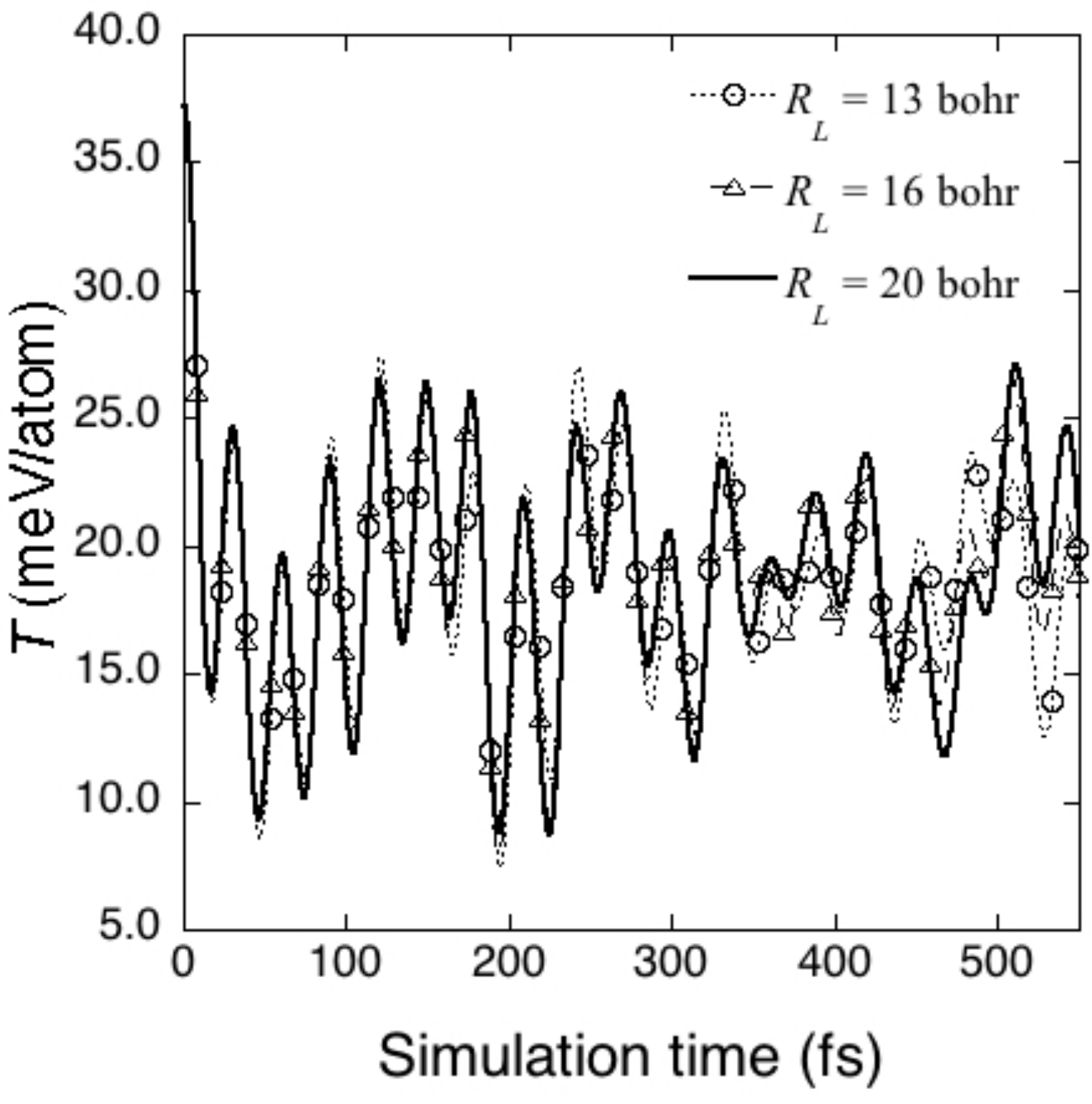}
                \caption{The ionic kinetic energy profiles by three $R_L$ values; 13 bohr, 16 bohr and 20 bohr.}
                \label{fig:TRLdep}
        \end{center}
\end{figure}

During MD, atoms will move in and out of the density matrix range, and we might expect to see
energy drift and deviations at small ranges, compared to long ranges.
The lack of drift is extremely encouraging, showing that the XL-BOMD approach is well-suited
both to orthogonal and non-orthogonal basis sets, and linear scaling approaches.
It is also encouraging that, for relatively modest ranges, good accuracy is seen.
We plot the kinetic energy of the ions in Fig.~\ref{fig:TRLdep}, which shows that, while there is a
deviation for 13 bohr range after 2-300 femtoseconds, the 16 bohr range profile is extremely close to
the 20 bohr range profile, and therefore this range can be used with confidence over at least 
1ps\cite{note3}. 


\subsubsection{\bf{$\epsilon_L$}-dependence}

\begin{figure}[H]
        \begin{center}
                \includegraphics[width=10cm,height=9cm]{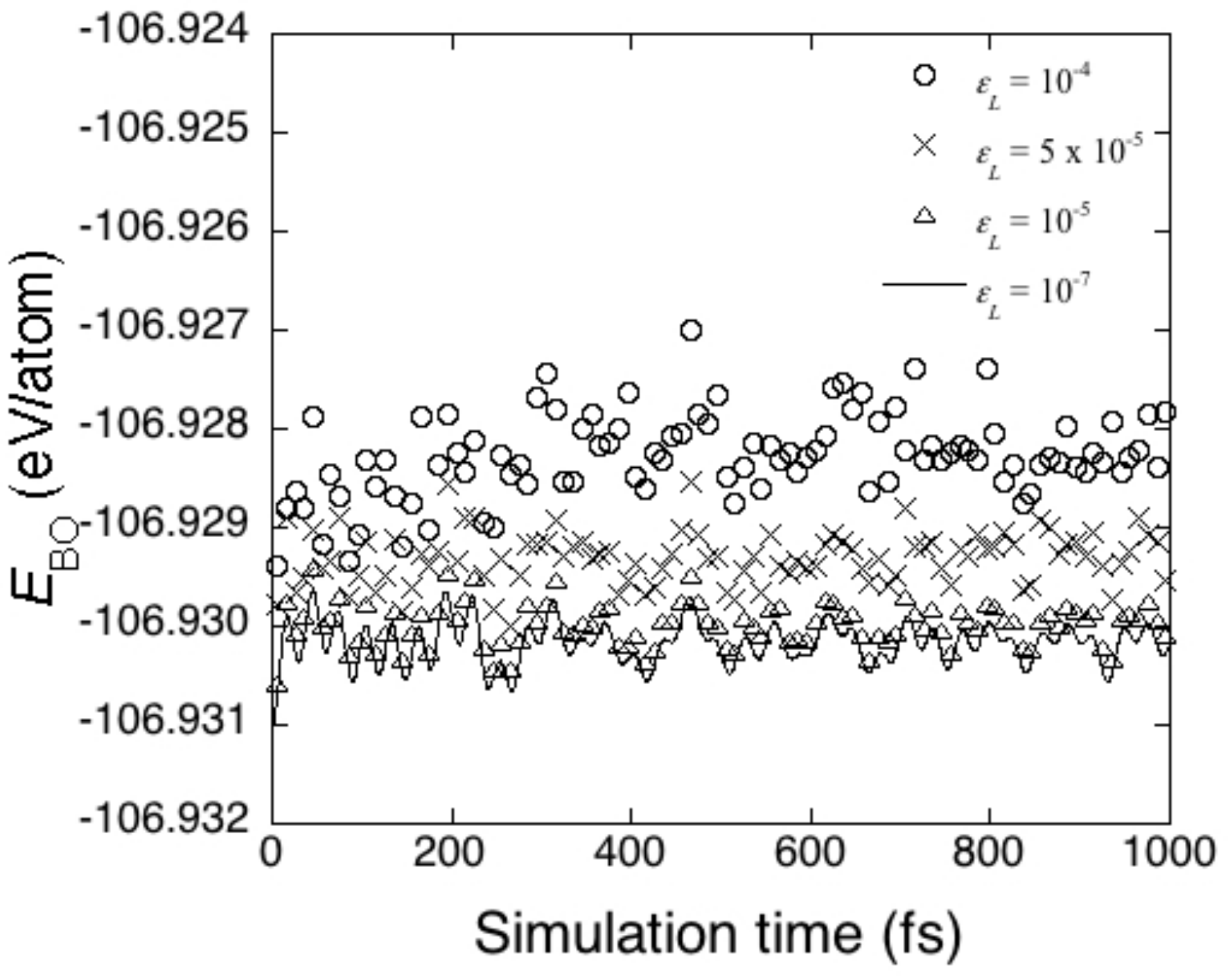}
                \caption{The Born-Oppenheimer total energy profiles calculated for different $\epsilon_L$ values: circles, squares and triangles denote the profiles given by $\epsilon_L=10^{-4}$,$5\times10^{-5}$ and $10^{-5}$, respectively. The solid line shows the profile calculated with $\epsilon_L=10^{-7}$.}
                \label{fig:E}
        \end{center}
\end{figure}

\begin{figure}[H]
        \begin{center}
                \includegraphics[width=10cm,height=9cm]{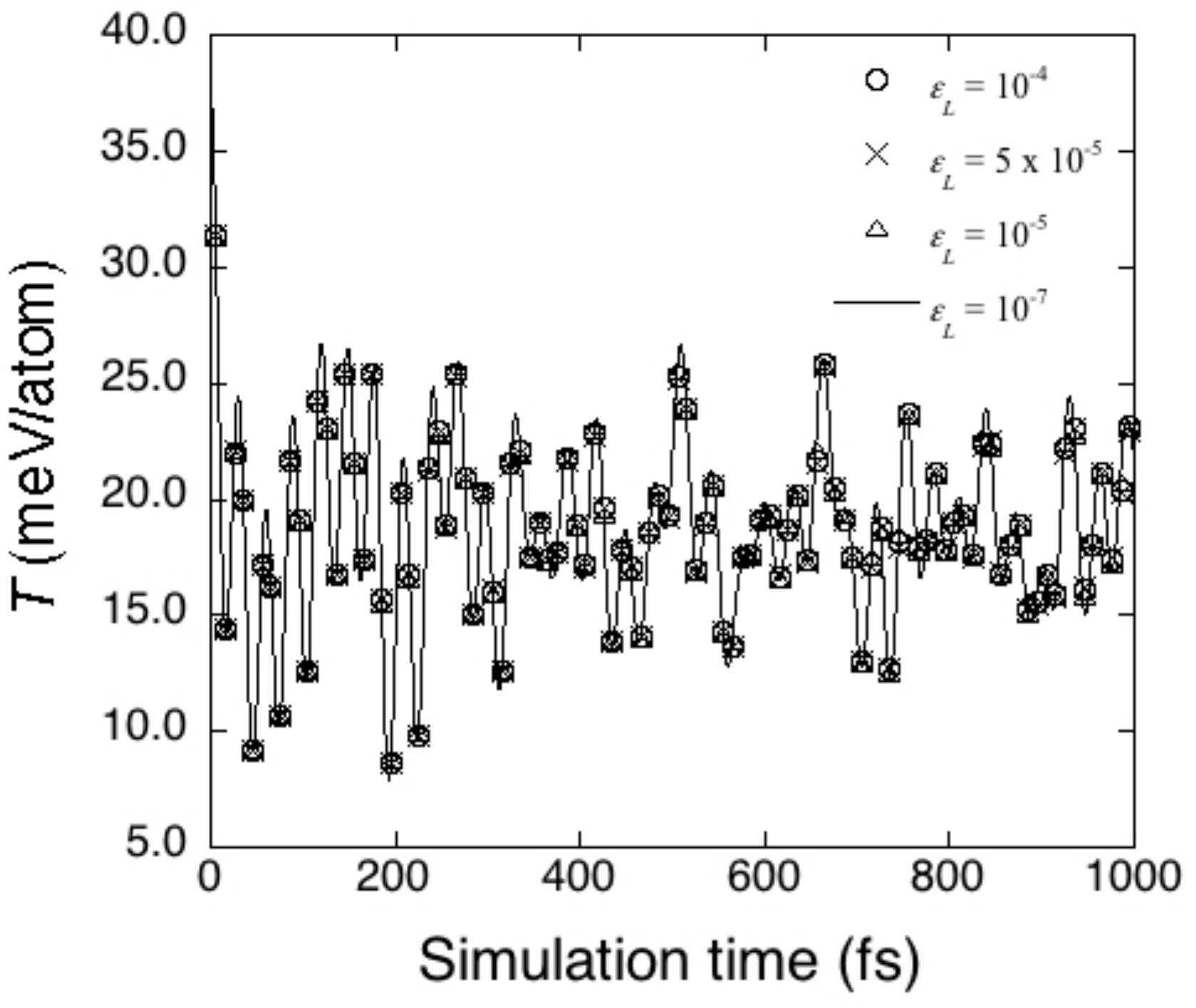}
                \caption{The ionic kinetic energy profiles calculated for different $\epsilon_L$ values: circles, crosses and triangles denote the profiles given by $\epsilon_L=10^{-4}$,$5\times10^{-5}$ and $10^{-5}$, respectively. A solid line shows the profile calculated with $\epsilon_L=10^{-7}$.}
                \label{fig:T}
        \end{center}
\end{figure}

The other main parameter which must be tested for its effect on the accuracy and stability
of linear-scaling MD is the DMM tolerance $\epsilon_L$.
From the results of the last section, we fix the $L$-range to 16 bohr and apply a wide range
of $\epsilon_L$ from $10^{-4}$ to $10^{-7}$; we assume that the results with $\epsilon_L=10^{-7}$
can be regarded as reference values.
We plot the BO total energy over a set of 1ps simulations in Fig.~\ref{fig:E}.
First we note that $E_{BO}$ with $\epsilon_L = 10^{-4}$ (circles) is not fully converged,
and is higher in energy by about 2 meV/atom than the converged one.
However, as we see in Fig.~\ref{fig:T}, the kinetic energy profile shows almost perfect agreement 
with the reference profile of $\epsilon_L = 10^{-7}$, indicating that forces and hence the ionic kinetic 
energy converge faster than the total energy.
Note that we can achieve good convergence for both $E_{BO}$ and $T$ with $\epsilon_L = 10^{-5}$.
In addition, the $E_{BO}$ profiles are close to each other for all tolerances except for a constant
shift of the energy.
Given the fast convergence in ionic kinetic energy, and the similarities in $E_{BO}$ profiles,
the calculated trajectories will be very close.
Hence, the DMM tolerance of $10^{-4}$ will give an accurate simulation.


\subsection{Practical \bf{$\mathcal{O}(N)$}-MD simulations}
\label{sec:pract-bmmathc-md}

Having investigated the effects of the parameters on simulations, we now present applications
of the \CQ\ code to practical MD simulations using a double-$\zeta$ + polarisation (DZP) PAO basis, 
which is commonly used as a converged basis set.
Three different systems are presented: 8-atom crystalline silicon; 32 water molecules;
and a 32,768-atom crystalline silicon cell to demonstrate application to a large system.
For all systems, $L$ range is set to 16 bohr. 
The initial temperature is 300K and the
time step of the simulation is 0.5 femtosecond, unless stated otherwise.

\subsubsection{Eight atom bulk silicon}
\label{sec:eight-atom-bulk}

\begin{figure}[H]
        \begin{center}
                \includegraphics[width=10cm,height=9cm]{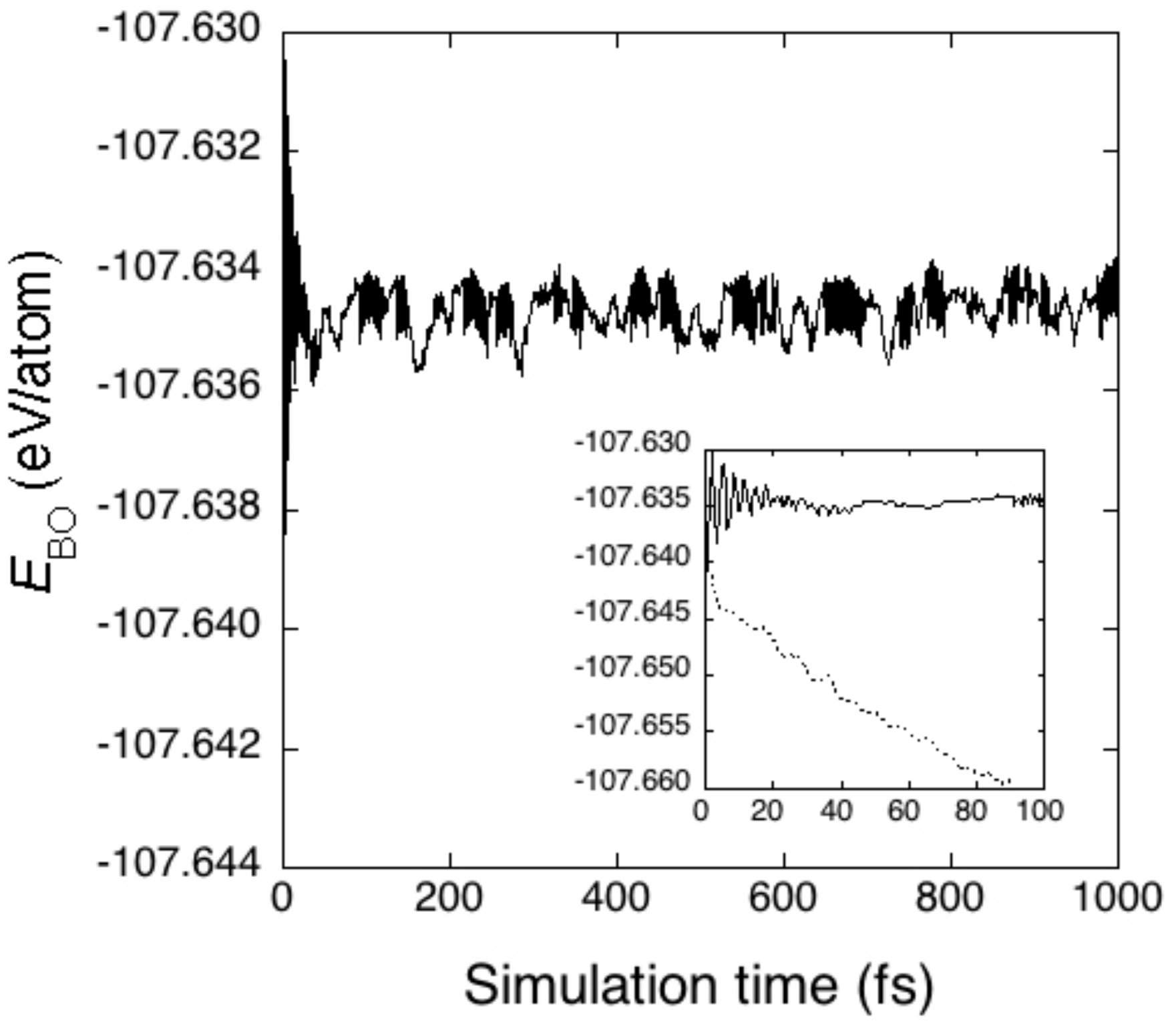}
                \caption{Main: Born-Oppenheimer total energy profile of crystalline Si with a DZP basis set.  
                 Inset: comparison of two approaches to XL-BOMD, solid and dotted lines indicating the 
                 corrected and non-corrected propagators, respectively.}
                \label{fig:BOESi8DZP}
        \end{center}
\end{figure}

Our first example is a small sample of crystalline silicon. We show the BO energy, $E_{\rm BO}$,
calculated with $\epsilon_L$ set to $1.3\times10^{-4}$ in Figure ~\ref{fig:BOESi8DZP}.
This plot demonstrates excellent energy conservation, though there are large fluctuations
for the first few femtoseconds.
This initial large amplitude is probably caused by the system not
being in an equilibrium state at the start of the simulation, and is not a significant problem.
In the inset of Fig. ~\ref{fig:BOESi8DZP}, we also show the result of a simulation
where we set $S =1$ in Eq.~\ref{eq:XLBOMD} or Eq.~\ref{eq:TRBOMD} (by supposing that $L$ is expressed using orthogonal basis sets).
By comparing the result (dotted line) with the correct one (solid line), we see that the
former energy exhibits significant drift, while the latter one is well behaved.
This indicates the importance of the metric used for non-orthogonal basis functions.


We performed another MD simulation on the same system using the same calculation conditions
except that the time step is increased to 2.0 femtoseconds (this result is shown in Supporting Information).
We see that the large oscillations observed in the early stage are even larger and the 
time to suppress them becomes longer.
However, it is found that the time evolution of the kinetic energy of Si atoms is almost the same
between these two MD simulations. 
This result means that time step of 2.0 femtoseconds can be used to do reliable MD simulations
on this silicon system.
It is encouraging that we can use such a large time step for O($N$) FPMD simulations. 
Here, we should note that a new efficient method which does not require self-consistent field optimization  
has been recently proposed\cite{2014.Souvatzis.JCP} within the XL-BOMD framework.
Introducing such new techniques to the {\sc Conquest} code may even further accelerate the speed of O($N$) FPMD simulations.

\subsubsection{Box of 32 water molecules}
\label{sec:box-water-bmtimes}

\begin{figure}[H]
        \begin{center}
                \includegraphics[width=10cm,height=9cm]{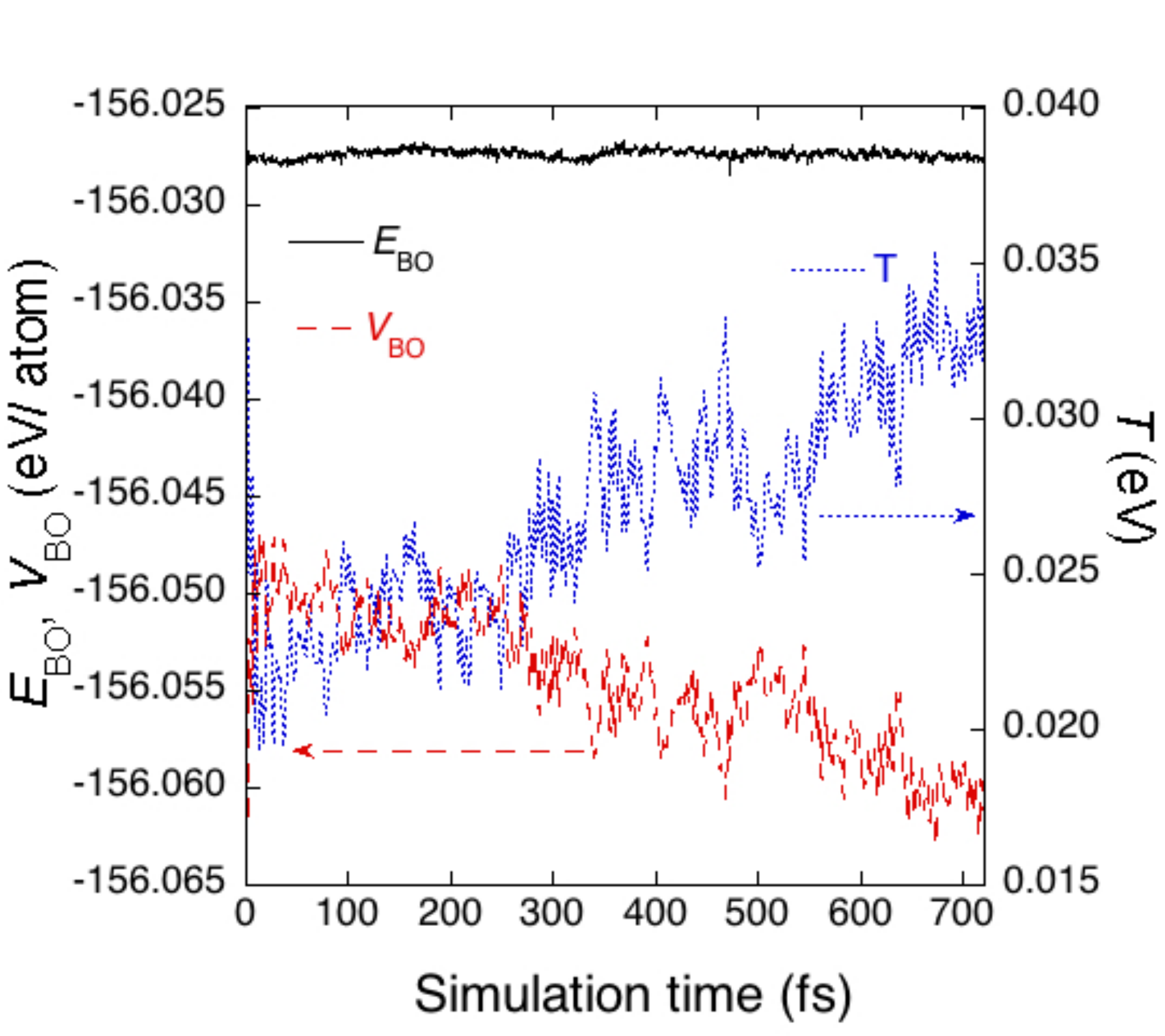}
                \caption{The Born-Oppenheimer total energy profiles of the waterbox system with DZP and PBE: 
                 a solid line represents $E_{\rm BO}$, and red dashed and blue dotted lines characterise 
                 its components, $V_{BO}$ and $T$, respectively (color online).}
                \label{fig:BOEwater32}
        \end{center}
\end{figure}

We take a box of water made up of 32 molecules as our next target.
Water is important as a system in its own right, as well as forming the environment for most biological simulations.
We performed linear scaling XL-BOMD with $\epsilon_L$ set to $10^{-4}$, and using the
GGA-PBE functional\cite{1996.perdew.3865,Torralba:2009JCTC1499} for
exchange-correlation, as it describes water better than LDA (we note
that the convergence with respect to the grid may be slower for GGA).
In Fig. ~\ref{fig:BOEwater32} we show the total energy and the ionic and potential energies, $T$ and $V_{\rm BO}$.
We see that there are no initial large oscillations unlike the silicon case, and that
$T$ appears to keep increasing during the run.
This behavior is seen because we used the micro-canonical ensemble
and the simulation time is too short to reach a steady state.
Even in such a situation showing fast changes, the averaged fluctuations are $\sim 1.1$ meV atom$^{-1}$ 
and good energy conservation is found without any energy drift.
These results clearly demonstrate that there are no effects from the
grid on the stability and accuracy of MD.

\subsubsection{32,768-atom crystalline silicon}
\label{sec:si-bmtimes-32}

\begin{figure}[H]
        \begin{center}
                \includegraphics[width=10cm,height=9cm]{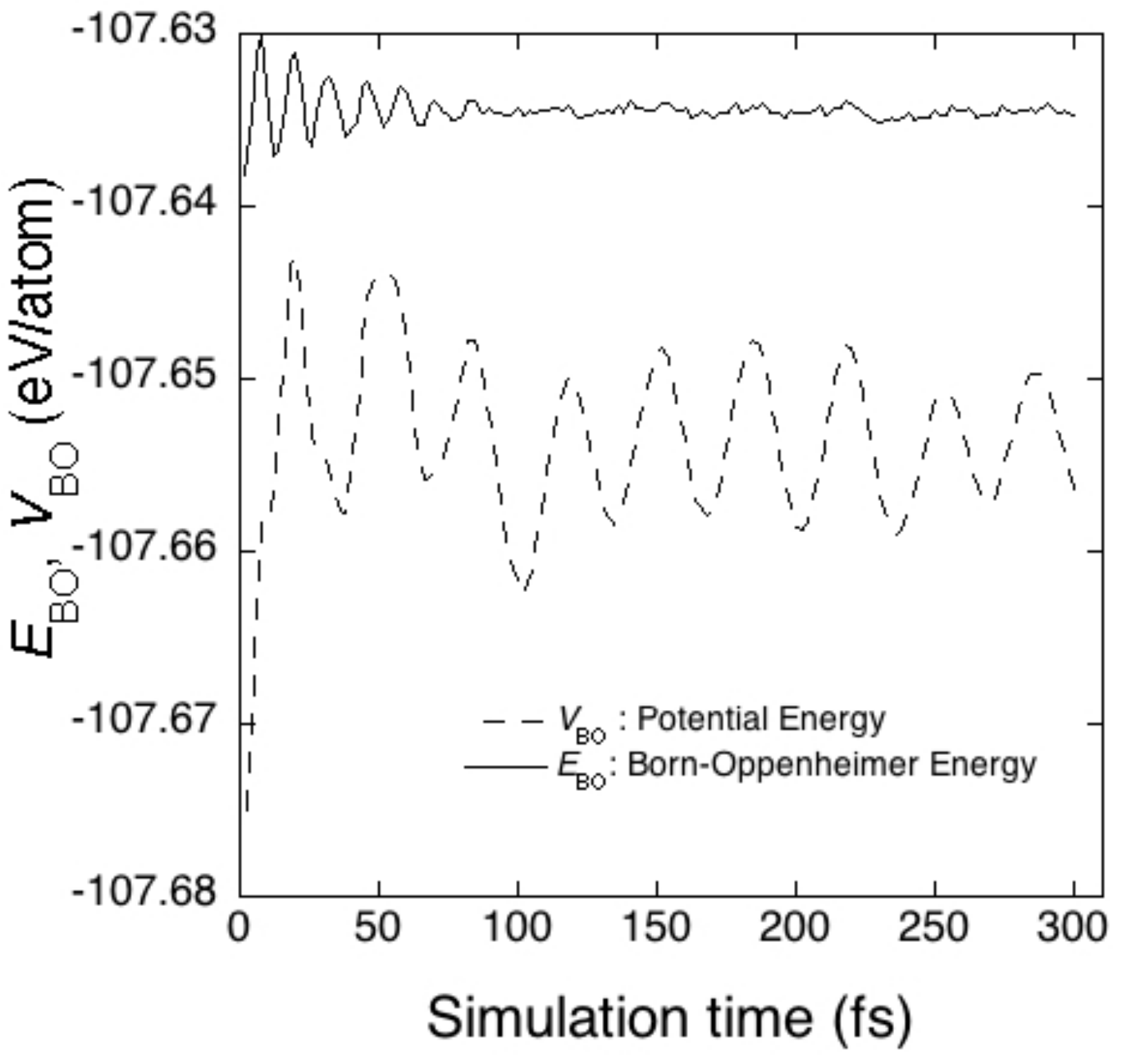}
                \caption{Profiles of the Born-Oppenheimer total energy and potential energy of a 32,768-atom crystalline 
                    silicon system with a DZP basis set and LDA functional.}
                \label{fig:Si32768}
        \end{center}
\end{figure}

Finally, as a demonstration of the scalability of our approach, we
apply our algorithm to a large system: a 32,768 atom crystalline silicon sample. 
The time evolution of the BO and potential energy are shown in Fig.~\ref{fig:Si32768}.
This is our first substantial attempt to examine the stability of an MD simulation on massive systems. 
This simulation is conducted under the same conditions as those in the 8-atom system 
in Fig.~\ref{fig:BOESi8DZP}, except that the time step is set to 2.0 femtoseconds.

We observe some similarities to the previous results. In the early stage of a simulation, 
large oscillations are seen in the energy, but are gradually suppressed during the MD run. 
The time to suppress the large oscillations is longer than the one in Fig.~\ref{fig:BOESi8DZP}, 
but similar to the case using 2.0 femtoseconds for 8-atom crystalline silicon.
(See Fig.~S1 in the Supporting Information.)
Thus, this comes from the difference of the time step, not from the system size.
The energy fluctuations after 100 femtoseconds is smaller than 1 meV atom$^{-1}$ 
and there is no energy drift, as shown in Fig.~\ref{fig:Si32768}. 
Although we only carried out a short MD run, the simulation is expected to remain stable 
in longer cases without any drift, based on the preceding example of an 8-atom system. 
Moreover, our approach is expected to ensure great stability on even larger and more complex systems 
by virtue of high computational efficiency realised in the \CQ\ 
code\cite{Bowler:2010uq,Arita.2014.JASSE,Miyazaki.2013.NIMSNOW}.
The average time required for 1 MD step was 1085 seconds using 1024
CPUs (with 16 cores/CPU) of Fujitsu FX10.
We are confident that this wall-clock time will be greatly reduced in
the near future, because the code is not optimized at present,
especially for the part updating $X$ matrices; for simplicity, we now
use a simple disk-based I/O procedure to
read and write matrix elements of $X$s in the previous steps and this part will be improved in the near future.



\section{Conclusions}

In this paper, we have investigated the efficiency, stability and accuracy of MD simulations with the
\CQ\ code using the DMM linear scaling approach, to demonstrate large-scale O($N$) FPMD simulations in practice.
We first found that accurate MD simulations can be performed by calculating the electronic structure from scratch
using the McWeeny procedure at every atomic configuration.  
This gives excellent energy conservation even with a rather rough tolerance on
the convergence of the density matrix minimisation.
However, this method requires considerable computational time because
it initialises the density matrix at each step, rather than re-utilising it.

Direct re-use of the optimised density matrix reduces computational
costs considerably, but causes an unphysical energy drift.
In order to achieve both accuracy and efficiency, we have introduced the XL-BOMD formalism, proposed by Niklasson
{\rm et al.}, to the \CQ\ code, formulated for non-orthogonal basis functions.
The application of XL-BOMD turned out to be much more efficient than
the McWeeny approach, especially with the aid of a dissipative force,
and good energy conservation was observed.
We studied how calculated results depend on two parameters specific to
the DMM method: the cutoff range of the $L$ matrix and the tolerance
applied to the density matrix minimisation.
We found that larger $L$-ranges resulted in smaller energy fluctuations.
Moreover, we found that the MD trajectories from runs with different ranges are almost identical, 
even when the Born-Oppenhimer total energy was not fully converged. This indicates that, as is well
known, the atomic forces converge much faster than the total energy
with respect to these two parameters.

As practical applications, we treated crystalline silicon with LDA and liquid water with GGA.
For the crystalline silicon, we demonstrated that accurate MD runs can be 
performed 
on very large systems, in this case containing 32,768 atoms, but scalable to significantly larger systems.
The algorithms presented in this paper are applicable to a canonical ensemble and thus will open doors
to more practical calculations on very large and complex systems.

\begin{acknowledgement}

The authors thank Lianheng Tong and Takao Otsuka for fruitful discussions.
This work is partly supported by KAKENHI projects by MEXT (No. 22104005)
and JSPS (No. 26610120 and 26246021), Japan. The support from the Strategic Programs 
for Innovative Research (SPIRE) and the Computational Materials Science Initiative (CMSI)
is also acknowledged. 
Calculations were performed using the Numerical Materials Simulator at NIMS, Tsukuba, Japan,
K computer at RIKEN Advanced Institute for Computational Science(AICS), Kobe, Japan,
and the Fujitsu FX10 system at the Information Technology Center, University of Tokyo,
Tokyo, Japan.

%

\end{acknowledgement}

\begin{suppinfo}

We present the comparison of the profiles of Born-Oppenheimer and kinetic energy
using different timesteps, 0.5 fs and 2.0 fs and show that the trajectory of the
ions are almost the same.
The result of crystalline silicon made of 4096 atoms
using the initial temperature of 600K is also provided and we see that the
energy convergence is excellent.

\end{suppinfo}


\begin{mcitethebibliography}{40}
\bibitem[foo()]{footnote1}
As each support function is represented by only one PAO, we need not consider
  the time evolution of $S$.\relax
\mciteBstWouldAddEndPunctfalse
\mciteSetBstMidEndSepPunct{\mcitedefaultmidpunct}
{}{\mcitedefaultseppunct}\relax
\EndOfBibitem
\bibitem[foo()]{footnote2}
This does not affect the results and larger basis sets are used below\relax
\mciteBstWouldAddEndPuncttrue
\mciteSetBstMidEndSepPunct{\mcitedefaultmidpunct}
{\mcitedefaultendpunct}{\mcitedefaultseppunct}\relax
\EndOfBibitem
\bibitem[not()]{note3}
This does not mean the result of MD by using $R_{L}$=13bohr is incorrect. The
  different trajectory may simply come from infinitesimal differences in forces
  of a chaotic system.\relax
\mciteBstWouldAddEndPunctfalse
\mciteSetBstMidEndSepPunct{\mcitedefaultmidpunct}
{}{\mcitedefaultseppunct}\relax
\EndOfBibitem
\end{mcitethebibliography}

\providecommand{\latin}[1]{#1}
\providecommand*\mcitethebibliography{\thebibliography}
\csname @ifundefined\endcsname{endmcitethebibliography}
  {\let\endmcitethebibliography\endthebibliography}{}

\end{document}